\documentstyle[11pt,aaspp4]{article}

\begin{document}

\title{The Inverse Compton Emission Spectra in the Very Early Afterglows 
of Gamma-Ray Bursts}

\author{X. Y. Wang,  Z. G. Dai  and T. Lu}
\affil{Department of Astronomy,
Nanjing University, Nanjing 210093, P.R.China
}
\affil{email:xywang@nju.edu.cn; daizigao@public1.ptt.js.cn; tlu@nju.edu.cn}

\begin{abstract}
We  calculate the spectra of inverse Compton (IC) emissions in gamma-ray
burst (GRB) shocks produced when relativistic ejecta encounters the external 
{ interstellar} medium,
assuming a broken power-law approximation to the synchrotron seed spectrum.
Four IC processes, including the synchrotron self-Compton (SSC)
 processes in GRB forward
and reverse shocks, 
and two combined-IC processes (i.e.
scattering of reverse shock photons on the electrons 
in  forward shocks and forward shock photons on the electrons in reverse shocks),
are considered.
We find that the SSC emission from  reverse shocks dominates over other emission processes
in  energy bands from tens of MeV to tens of GeV,
 for a wide range of shock parameters.
 This mechanism may be responsible for the prompt high energy gamma-rays detected
by the Energetic Gamma Ray Experiment Telescope (EGRET).
{ At TeV energy bands, however, the combined-IC emissions 
and/or the SSC emission from the forward shocks 
become increasingly dominant for a moderately steep  distribution of shocked
electrons.}
 \end{abstract}


\section{Introduction}
The current standard model for gamma-ray bursts (GRBs) and their afterglows
is the fireball shock model (see Piran 1999 for a recent review).
It involves  a large amount of  isotropic equivalent energy release, 
$E_{0} \sim10^{52-54}$ ergs,
within a few seconds and in a small volume with negligible
baryonic load, which leads to a fireball that expands
ultra-relativistically  into the external 
medium. 
A substantial fraction of the kinetic energy of the baryons is transferred
to a non-thermal population of relativistic electrons through Fermi
acceleration in the shock (e.g. M\'esz\'aros $\&$
Rees 1993). The accelerated electrons cool via synchrotron emission and
inverse Compton scattering in the post-shock magnetic fields and produce
the radiation observed in GRBs and their afterglows (e.g.
Paczy\'{n}ski \& Rhoads 1993; Katz 1994;
Sari et al. 1996; Vietri 1997; Waxman 1997a; Wijers et al. 1997).
The shock could be either $\it{internal}$ due to collisions between
fireball shells caused by outflow variability (Paczy\'{n}ski
$\&$ Xu 1994; Rees $\&$ M\'esz\'aros 1994),
or $\it{external}$ due to the interaction of the fireball with the
surrounding interstellar or wind media ( M\'esz\'aros
$\&$ Rees 1993; Dai \& Lu 1998; 
Chevalier \& Li 1999). 

When the relativistic ejecta encounters the external medium,
 a relativistic {\em forward} shock expands into the external medium and a {\em reverse}
 shock moves into and heats the fireball ejecta. The shocked ambient and 
 ejecta materials are in pressure balance and separated by a contact discontinuity.
The forward shock continuously heats fresh gas and accelerates electrons,
producing long-term afterglows through the synchrotron emission
(e.g. Waxman 1997a,b; Wijers et al.
1997; Vietri 1997; Wang et al. 2000a,b;  Dai \& Lu 1999, 2000; Dai et al. 1999; 
Huang et al. 1998, 2000; Gou et al. 2001). On the other hand, 
the reverse shock operates only once and after that 
emission from the fireball ejecta are suppressed after the reverse shock  
crosses the ejecta, since then the ejecta expands and cools adiabatically.

When a reverse shock crosses the shell, the  shocked shell and 
the forward shocked external medium 
 carry comparable amount of energy. However,
the typical temperature of the shocked shell electrons is lower since the
particle number density is higher. So, the typical frequency of the synchrotron
radiation from the shocked shell is considerably lower. A strong prompt optical
flash (Akerlof et al. 1999) and late time radio flare behavior (Kulkarni et al.
1999),  accompanying GRB990123,
have been attributed to the synchrotron emission process from this 
reverse shock (Sari \& Piran 1999; {M\'esz\'aros} \& Rees 1999). 

In a recent paper (Wang, Dai \& Lu 2001; hereafter WDL), we analytically study the 
high energy $\gamma$-ray emission from the SSC process in the reverse shocks.
The result showed that this emission dominates over the components of the 
the synchrotron and IC emissions from  forward shocks at high energy $\gamma$-ray
bands from tens of Mev to tens of GeV. Furthermore, we suggest that 
this mechanism may explain 
the observations of the prompt high energy gamma-rays detected by EGRET, such as 
those from
GRB930131.

In this work, we numerically calculate the SSC radiation components in the forward
and reverse shocks. We
also consider another two combined-IC processes, i.e. 
scatterings of reverse shock photons on the forward  shocked electrons 
 and forward shock photons on the reversely  shocked electrons.
 For a wide range of shock parameters, the present numerical
 results confirm  our previous suggestions that the SSC emission from
 reverse shocks is  the most important at the energy bands to which EGRET
is sensitive.

While  IC emissions from  afterglow forward shocks 
(e.g. Sari et al. 1996; { Totani 1998a;} Waxman 1997; Panaitescu \& M\'esz\'aros
1998; Wei \& Lu 1998, 2000; Chiang \& Dermer 1999; 
{ Dermer, B${\rm {\ddot o}}$ttcher \& Chiang 2000; Dermer, Chiang \& Mitman 2000;}
Panaitescu \& Kumar 2000)
 and GRB internal shocks  
 (Papathanassiou \& M\'esz\'aros 1996; Pilla \& Loeb 1998; Panaitescu \& M\'esz\'aros
2000) have attracted a lot of investigation, the studies for IC process
 in reverse shocks  and 
combined-IC scatterings between reverse and forward shocks 
(i.e.  very early afterglows) are still quite  preliminary (see e.g. M\'esz\'aros, Rees \&
Papathanassiou 1994).

In section 2, we briefly describe the hydrodynamics evolution of a fireball shell.
We present in section 3 the synchrotron seed spectra and 
shocked electron distributions,  then 
perform a numerical calculation of IC emissions in section 4. Finally,
we give a summary and discussion.

\section{Hydrodynamics of a fireball shell}
Let's consider an ultrarelativistic cold shell with energy $E$, initial Lorentz factor
$\eta$ and a width $\Delta$ in the observer frame  expanding into a cold 
external { interstellar medium (ISM)}.
When the shell sweeps up a large volume of ISM matter, it begins to be decelerated
significantly. The interaction between the shell and the ISM matter
 leads to two shocks:
a forward shock propagating into the { ISM } and a reverse shock moving back into
the shell. There are four regions separated by the two shocks and a contact
discontinuity: the cold { ISM }(denoted by the subscript 1), the shocked
 ISM (2),
the shocked shell material (3) and the unshocked shell material (4).
From the shock jump conditions and the equalities of pressure and velocity
along the contact discontinuity, we can evaluate the Lorentz factor $\gamma$,
the pressure $p$ and the number density $n$ in the shocked regions in terms of 
three variables $n_1$, $n_4$ and $\eta$ (Blandford \& McKee 1976; Sari \&
Piran 1995).

Sari \& Piran (1995) showed that, if the shell is thin: $\Delta<l/2\eta^{8/3}$,
where $l\equiv(3E/4\pi{n_1}m_p{c^2})^{1/3}$ is the Sedov length,
 the reverse shock
is Newtonian (with shell spreading, the reverse shock can be mildly relativistic),
which means that the Lorentz factor of the shocked shell material $\bar{\gamma}_3$
is almost unity in the frame of the unshocked shell and ${\gamma}_3\sim\eta$.
On the other hand, if the shell is thick: $\Delta>l/2\eta^{8/3}$, the reverse
shock is relativistic with $\bar{\gamma}_3\sim
\eta/2\gamma_3$ and it then considerably decelerates the shell material (Kobayashi 2000).

{ After the shell being decelerated, the evolution of the adiabatic 
forward shock follows the Blandford-McKee solution (Blandford \& McKee 1976).
For the shocked shell, its  dynamic evolution may be complex. 
Kobayashi \& Sari (2000) showed that if the fireball shell has relativistic
temperature, the Blandford-McKee solution can be regarded as an adequate description
quite early on. But for the case that the reverse shock is only mildly-relativistic,
the evolution of the shocked shell may deviate the Blandford-McKee solution.
This may affect the time evolution of the late-time reverse shock emission, but
does not affect our computation of the peak-time emission. }

\section{Synchrotron emission spectra and electron distributions in GRB shocks}

\subsection{The synchrotron seed spectrum}

The IC emissions depend on both the seed photon spectrum and the 
shocked electron distribution.
We assume the  synchrotron seed spectrum to be 
described by four broken power-law segments as given in Sari et al. (1998).
The distributions of newly-shocked electrons in both the forward and reverse shocks
are assumed to be a power law of index $p$ ($N(\gamma)\propto{\gamma}^{-p}$),
{ with the minimum Lorentz factor of shocked electrons in the shell 
rest frame being $\gamma_{m}=\frac{m_p}{m_e}\frac{p-2}{p-1}\xi_{e}{\gamma_{sh}}$,
where $\xi_{e}$ is the fraction of thermal energy carried by electrons
(Sari, Piran \& Narayan 1998)}. Here for the forward shock 
$\gamma_{sh}=\gamma_2=\gamma_3\equiv{\Gamma}$ 
and $\gamma_{sh}=\bar{\gamma}_3$ for the reverse shock.
Assuming that $\xi_B$ is the fraction of the thermal energy carried by 
the magnetic field, the magnetic field strength is
 $B'=12{\rm G}{\xi_{B,-2}}^{1/2}({\gamma_2}/{300})n_{1,0}^{1/2}$,
 where $\xi_{B,-2}=\xi_B/10^{-2}$, and $n_{1,0}=n_1/10^0$.
 The shocked relativistic electrons cool through synchrotron emission and 
 IC scatterings of the synchrotron photons; thus the Lorentz factor of the 
 electrons that cool on a timescale equal to the dynamic timescale is 
 given by
 \begin{equation}
 \gamma_c=\frac{6\pi{m_e}c(1+z)}
{(Y+1)\sigma_T\gamma_2(B')^2{t}}=
 \frac{4500}{Y+1}\xi_{B,-2}^{-1}
(\frac{\gamma_2}{300})^{-3}t_1^{-1}(\frac{1+z}{2})n_{1,0}^{-1},
\end {equation}
where $t_1=t/10\,{\rm s}$ denotes the time in the observer frame, $Y$ is the Compton
parameter, $z$ is the redshift of the burst source and $\sigma_T$ is the Thomson
scattering cross section.
{ The Compton parameter $Y$, defined to be the ratio 
between the inverse Compton luminosity of electrons and the synchrotron luminosity,
 can be expressed as $Y=L_{IC}/L_{syn}=U_{rad}/U_{B}
\sim(-1+\sqrt{1+4\eta_e\xi_e/\xi_B})/2$, where 
$U_{syn}$ and $U_{B}$ are the energy density of synchrotron radiation and
magnetic field , respectively, and
  $\eta_e$ represents
the fraction of the electron energy that was radiated away ( Sari \& Esin 2001;
Panatescu \& Kumar 2000). For fast cooling electrons,
$Y$ is estimated to $\sqrt{\xi_e/\xi_B}$ and it may be smaller for slow
cooling electrons. However, as the two combined-IC processes 
(i.e. scattering of reverse shock photons on the electrons 
in  forward shocks and forward shock photons on the electrons in reverse shocks)
are considered here, the IC cooling of the electrons in both shocks should
be dominated by the larger of the two photon fields. Because both shocks have 
comparable energy, the magnetic filed energies in the two shocks are comparable,
and therefore $Y$ in two shocks is almost the same and equals to
 $\sqrt{\xi_e/\xi_B}$ as long as one shock is in the fast cooling regime. }

According to the shell thickness, we divided our discussion into two cases:
the thin shell case and the thick shell case. The thin shell case implies that
the deceleration timescale $t_{dec}$ (defined as the observer time at which the 
heated ISM energy is comparable to the initial energy $E$), i.e.
 \begin{equation}
{t_{dec}}=\frac{r_{dec}}{2\eta^2{c}}(1+z)=10  (\frac{1+z}{2}){E_{53}^{1/3}}
{n_{1,0}^{-1/3}}{\eta_{300}^{-8/3}}\,{\rm s},
\end{equation}
is larger than the shell crossing time ($\Delta/c$), 
where $E_{53}=E/10^{53}{\rm erg}$ and $\eta_{300}=\eta/300$,
while the thick shell
has $\Delta/c>t_{dec}$.
The reverse shock emission peaks at $t_{dec}$ for the thin shell case and at 
$\Delta/c$ for the thick shell. Our following calculations
 of the IC spectra correspond to this
peak time.

\subsubsection{ The thin shell case}

As described by Sari et al. (1998), the synchrotron radiation from the shocked electrons
can be approximated by a broken power-law spectrum with three characteristic break
frequencies. One is the self-absorption frequency, $\nu_a$. The other two are the
peak frequencies of the emission from the electrons with the characteristic Lorentz
factor $\gamma_m$ and the cooling Lorentz factor $\gamma_c$, denoted as $\nu_m$ and $\nu_c$, respectively.
{  
As usual,  the fraction of thermal energy carried by electrons, $\xi_e$, and magnetic
field, $\xi_B$, is assumed to be similar in the forward and reverse shock. At the peak time
$t_{dec}$ of the reverse shock emission,  the break frequencies $\nu_m$ from the two shocks
are, respectively,
\begin{equation}
\nu_{m}^{rs}=\frac{1}{1+z}\frac{\Gamma(\gamma_m^{rs})^2 e B'}{2\pi m_e c}=
6.4\times10^{15}(\frac{p-2}{p-1})^2(\frac{\xi_{e}}{0.6})^2\xi_{B,-2}^{1/2}
\eta_{300}^2{n_{1,0}^{1/2}}(\frac{2}{1+z})~{\rm Hz},
\end{equation}
and
\begin{equation}
\nu_m^{fs}=\frac{1}{1+z}\frac{\Gamma(\gamma_m^{fs})^2 e B'}{2\pi m_e c}=
5.6\times10^{20}(\frac{p-2}{p-1})^2(\frac{\xi_e}{0.6})^2\xi_{B,-2}^{1/2}
\eta_{300}^4{n_{1,0}^{1/2}}(\frac{2}{1+z}){~\rm Hz}.
\end{equation}
The equipartition values are chosen to be the ones inferred for GRB970508:
$\xi_e=0.6$ and $\xi_B=0.01$ (Granot et al. 1999; Wijers \& Galama 1999).
Similarly, the cooling break frequencies of both shocks are  given by
\begin{equation}
\nu_c^{rs}=\nu_c^{fs}=\frac{1}{1+z}\frac{\Gamma(\gamma_c)^2 e B'}{2\pi m_e c}
=\frac{10^{17}}{(Y+1)^2}E_{53}^{-1/2}\xi_{B,-2}^{-3/2}
{n_{1,0}^{-1}}(\frac{t_{dec}}{10\,\rm s})^{-1/2}(\frac{2}{1+z})^{1/2}~{\rm Hz}.
\end{equation}
The peak flux of the synchrotron radiation is given by (Wijers \& Galama 1999)
\begin{equation}
f_{m}=\frac{N_e \Gamma P'_{\nu_m}(1+z)}{4\pi D_L^2}
\end{equation}
where $P'_{\nu_m}$ is the peak spectral power and $D_L$ is the luminosity distance 
of the burst. Note that for the forward shock, $N_e$ is the total number of swept-up
ISM electrons while for the reverse one, $N_e$ is the total number of shocked shell
electrons. Then, we obtain the peak flux of the two shocks at the time $t_{dec}$:
\begin{equation}
f_m^{rs}=4.8\,(\frac{1+z}{2})D_{L,28}^{-2}\xi_{B,-2}
^{1/2}{\eta_{300}^{-1}}{n_{1,0}^{1/4}}{E_{53}^{5/4}}(\frac{2}{1+z}\frac
{{t_{dec}}}{10{\,\rm s}})^{-3/4}~{\rm Jy},
\end{equation}
\begin{equation}
f_m^{fs}=26\,(\frac{1+z}{2})D_{L,28}^{-2}
\xi_{B,-2}^{1/2}E_{53}{n_{1,0}^{1/2}}{\,\rm mJy},
\end{equation}
where $D_{L,28}=D_L/10^{28}{\rm cm}$.
}

Now we derive the synchrotron self-absorption frequency of the reverse shock emission.
In the comoving frame of the shocked gas (denoted with a prime), 
the absorption coefficient $\alpha'_{\nu'}$ scales as $\alpha'_{\nu'}\propto
{\nu'}^{-(p+4)/2}$ for ${\nu'}>\nu'_p\equiv
\min(\nu'_m, \nu'_c)$ and as $\alpha'_{\nu'}\propto
{\nu'}^{-5/3}$ for ${\nu'}<\nu'_p$. In this frame, the absorption coefficient
for ${\nu'}<\nu'_p$ is given by (Rybicki \& Lightman 1979)
\begin{equation}
\alpha'_{\nu'}=\frac{\sqrt{3}e^3}{8\pi m_e}\left(\frac{3e}{2\pi m_e^3c^5}
\right)^{p/2}(m_ec^2)^{p-1}KB'^{(p+2)/2}
\Gamma\left(\frac{3p+2}{12}\right)\Gamma\left(\frac{3p+22}{12}\right)
{\nu'_p}^{-(p+4)/2}\left(\frac{\nu'}{\nu'_p}\right)^{-5/3},
\end{equation}
where $K=(p-1)n_4(4{\bar\gamma}_3+3){ \gamma_p^{p-1}}$, ${ \gamma_p\equiv\min(\gamma_m,\gamma_c)}$,
 $e$ is the electron charge
and $\Gamma(x)$ is the Gamma function.
Noting that the shocked shell width in the comoving frame $\Delta{r'}=r/\eta(4
{\bar\gamma}_3+3)$ and setting 
\begin{equation}
\tau(\nu'_a)\equiv\alpha'_{\nu'_a}\Delta{r'}=1,
\end{equation}
we get the synchrotron self-absorption frequency $\nu_a^{rs}$ in the observer
frame,
\begin{equation}
\nu_a^{rs}={ 1.4\times10^{13}}E_{53}^{1/5}\eta_{300}^{8/5}
(\frac{\xi_e}{0.6})^{-1}\xi_{B,-2}^{1/5}n_{1,0}^{9/10}(\frac{2}{1+z})\, {\rm Hz}
\end{equation}
for $p=2.5$.
Similarly, one can obtain the  synchrotron self-absorption frequency
of the forward shock emission,
\begin{equation}
\nu_a^{fs}={ 2.2\times10^{11}}E_{53}^{7/10}\xi_{B,-2}^{6/5}n_{1,0}^{11/10}
(\frac{t_{dec}}{10\, {\rm s}})^{-1/2}(Y+1)(\frac{2}{1+z})^{1/2} \, {\rm Hz}.
\end{equation}
{
Please note that because the forward shock always locates in front of the reverse
shock, there will be almost no detected reverse shock emission below the self-absorption
frequency of the forward shock. Thus, the  synchrotron spectrum of the 
slow-cooling reverse shock 
 is  described by  }
\begin{equation}
 f_{\nu}^{rs} = f_m^{rs} \left\{ \begin{array}{llll}
  {  0    }                                       &{ \nu < \nu_a^{fs} }\\
  (\nu/\nu_a^{rs})^2 (\nu_a^{rs}/\nu_m^{rs})^{1/3}   & \nu_a^{fs}< \nu < \nu_a^{rs}  \\
  (\nu/\nu_m^{rs})^{1/3}                        &\nu_a^{rs}<\nu<\nu_m^{rs}\\
  (\nu/\nu_m^{rs})^{-(p-1)/2}                   &\nu_m^{rs}<\nu<\nu_c^{rs}\\
  (\nu/\nu_c^{rs})^{-p/2} (\nu_m^{rs}/\nu_c^{rs})^{(p-1)/2}  &\nu_c^{rs}<\nu 
                                             \end{array} \right. \;,  
\end{equation}
while for the forward shock, which is in the fast cooling regime for typical
shock parameters,
\begin{equation}
 f_{\nu}^{fs} =f_m^{fs}  \left\{ \begin{array}{lll}
  (\nu/\nu_a^{fs})^2 (\nu_a^{fs}/\nu_c^{fs})^{1/3} &\nu < \nu_a^{fs}  \\
  (\nu/\nu_c^{fs})^{1/3}                 &\nu_a^{fs}<\nu<\nu_c^{fs}\\
  (\nu/\nu_c^{fs})^{-1/2}                   &\nu_c^{fs}<\nu<\nu_m^{fs}\\
  (\nu/\nu_m^{fs})^{-p/2}(\nu_c^{fs}/\nu_m^{fs})^{1/2}  &  \nu_m^{fs}< \nu 
                             \end{array} \right.  \;.
\end{equation}

\subsubsection{The thick shell case}
In a thick shell case, the reverse shock becomes relativistic before it crosses
the entire shell and begins to decelerate the shell material. It crosses a thick
shell at $T\sim\Delta/c$ and the peak time of the reverse shock emission is 
comparable to the GRB duration $T$. The Lorentz factor of the shocked shell
scales with time as ({ Sari 1997})
\begin{equation}
\gamma_3=(\frac{l}{\Delta})^{3/8}(\frac{4ct}{\Delta})^{-1/4},
\end{equation}
thus the Lorentz factor of the reverse shock at this peak time is 
\begin{equation}
\bar{\gamma}_3=\eta/2{\gamma}_3={ 2\,}\eta_{500}E_{53}^{-1/8}
(\frac{2}{1+z}\frac{T}{{ 100}~{\rm s}})^{3/8}n_{1,0}^{1/8}.
\end{equation}
The reference values of the comoving width and the initial Lorentz factor of the thick
shell have been chosen to be  ${\Delta=3\times10^{12}{\rm cm}}$  and $ \eta=500$, 
respectively. The shock radius at the peak time of the reverse shock emission
is
\begin{equation}
r=2\gamma_3^2{cT}/(1+z)={{ 5 }\,\times10^{16}~E_{53}^{1/4}n_{1,0}^{-1/4}
(\frac{2}{1+z}\frac{T}{{ 100~}} \rm s})^{1/4}{~\rm cm}.
\end{equation}
Using the expressions of $\gamma_m$, $\gamma_c$, $Y$ parameter, 
and Eq.(7), we obtain
the break frequencies and peak flux of 
the reverse shock synchrotron spectrum:
\begin{equation}
\nu_{m}^{rs}=
{ 1.8\times10^{16}}(\frac{p-2}{p-1})^2(\frac{\xi_{e}}{0.6})^2\xi_{B,-2}^{1/2}
\eta_{500}^2{n_{1,0}^{1/2}}(\frac{2}{1+z})~{\rm Hz},
\end{equation}
\begin{equation}
\nu_c^{rs}=\frac{{ 7.6\times10^{15}}}{(Y+1)^2}\xi_{B,-2}^{-3/2}
E_{53}^{-1/2}{n_{1,0}^{-1}}(\frac{T}{{ 100}\rm s})^{-1/2}(\frac{2}{1+z})^{1/2}{~\rm Hz},
\end{equation}
\begin{equation}
f_m^{rs}={ 0.3} (\frac{1+z}{2})D_{L,28}^{-2}\xi_{B,-2}
^{1/2}{\eta_{500}^{-1}}{n_{1,0}^{1/4}}{E_{53}^{5/4}}(\frac{2}{1+z}\frac
{{T}}{100{\rm s}})^{-3/4}~{\rm Jy}.
\end{equation}
Noting that the shocked shell width in the comoving frame $\Delta{r'}=\Delta\gamma_2/(4
{\bar\gamma}_3+3)$ and the number density $n_3=(4{\bar\gamma}_3+3)
E/(4\pi{m_p}c^2{\eta\gamma_2}\Delta{r}^2)$, we can also obtain the self-absorption break
 frequency $\nu_a$
using Eq.(10)
for the thick shell case.
Since for a thick shell, the reverse shock is generally in the fast cooling regime,
the synchrotron spectrum is described in a  way similar to Eq.(14).

\subsection{Electron distribution in shocked material}
The electron distribution $N(\gamma)$ in shocked shell or shocked external medium
is determined by the initial shocked electron distribution $N_i(\gamma)$ and by
electron cooling effect through synchrotron and IC radiation and possibly by the 
self-absorption of the synchrotron photons. If the newly-shocked electrons with
typical Lorentz factor $\gamma_m$ cools faster than the shock dynamics timescale,
the resulting electron distribution takes the form 
\begin{equation}
N(\gamma)\propto\left\{
       \begin{array}{ll}
       \gamma^{-2}& {\rm if}\,\,\,\gamma_c<\gamma<\gamma_m \\
       \gamma^{-p-1}    & {\rm if}\,\,\,
                   \gamma_m<\gamma<\gamma_{max} \\
          
       \end{array}
       \right. ,
\end{equation}
where $\gamma_{max}$ is the maximum Lorentz factor of shocked electron, which is 
determined by equating the electron acceleration timescale with the synchrotron
cooling timescale (e.g. {M\'esz\'aros}, Laguna \& Rees 1993), i.e.
$\gamma_{max}=10^{8}B'^{-1/2}$.
In the opposite case, most electrons have a random Lorentz factor $\gamma_m$,
and the electron distribution is
\begin{equation}
N(\gamma)\propto\left\{
       \begin{array}{ll}
       \gamma^{-p}& {\rm if}\,\,\,\gamma_m<\gamma<\gamma_c \\
       \gamma^{-p-1}    & {\rm if}\,\,\,
                   \gamma_c<\gamma<\gamma_{max} \\
          
       \end{array}
       \right.
\end{equation}

\section{ IC emissions from very early external shocks}
After having obtained the seed photon spectrum and the electron distribution,
now we can compute the up-scattering emissions of the synchrotron radiation
by relativistic electrons in the very early external shocks. We consider only the
first-order IC and neglect higher order IC processes, because a once-scattered
typical photon by a typical electron with $\gamma_e$ has energy of order
$(h\nu)_{\rm com}\gamma_e^3\ga{m_e{c^2}}$ in the rest frame of the second scattering 
electron. Then we can no longer use
Thomson limit to the scattering cross section, and the energy gained of the scattered
photon in each successive scattering will be reduced due to electron recoil and
to the necessity of using Klein-Nishina scattering cross section (Sari \& Esin 2001).

For single scattering, the IC volume emissivity in the comoving frame for a distribution
$N(\gamma) $ of scattering electrons is given by (Rybicki \& Lightman 1979;
Sari \& Esin 2001)
\begin{equation}
j_{\nu'}^{'IC}=3\sigma_T\int_{\gamma_{min}}^{\gamma_{max}}d\gamma{N(\gamma)}
\int_0^1{dx}{g(x)}\bar{f}'_{\nu'_s}(x),
\end{equation}
where $x\equiv\nu'/4\gamma^2{\nu'_s}$, 
$\bar{f}'_{\nu'_s}$ is the incident specific flux at the shock front in the 
comoving frame, and $g(x)=1+x+2x{\rm ln}(x)-2x^2$ reflects the angular dependence
of the scattering cross section for $\gamma_e\gg1$ (Blumenthal \& Gould 1970).
Noting that $f_{\nu'}^{'IC}=j_{\nu'}^{'IC}4\pi{r^2}\Delta{r'}/4\pi{D^2}$ and
the synchrotron flux 
$f'_{\nu'}=\bar{f}'_{\nu'_s}4\pi{r^2}/4\pi{D^2}$, where $\Delta{r'}$ is the comoving 
width of the shocked shell or ISM medium and $D$ is source distance, 
we obtain the IC flux in the observer frame
\begin{equation}
f_{\nu}^{IC}=3\Delta{r'}\sigma_T\int_{\gamma_{min}}^{\gamma_{max}}d\gamma{N(\gamma)}
\int_0^1{dx}{g(x)}f_{\nu}(x)
\end{equation}
by transforming Eq.(23) into the observer frame.

Apart from the SSC scattering processes in the reverse and forward shocks,
another two combined-IC scattering processes
 are also present. 
 Because approximately one-half of the photons
 arised in one shock region will diffuse into the another shock
 region from the point of view of the comoving frame, the IC flux Eq.(24) for
 the combined-IC scatterings 
  should be 
 divided by a factor of two.  Though  the scattered photons
 move isotropically in the comoving frame, the beaming effect makes these photons
 moving along the direction to the observer.

Our main calculation results are as follows:

i) The IC spectral flux from a relativistic thin shell 
expanding into  an ISM at the deceleration
time are shown in Fig. 1. Typical shock parameters are used: 
$E=10^{53}{\rm erg}$, { $\xi_e=0.6$}, $\xi_B=0.01$, $p=2.5$ and $n_1=1$.
Four IC spectra are displayed in the figure, including two SSC spectra
 from the reverse shock
({\em solid curve}) and forward shock ({\em dotted curve}), respectively, 
scatterings of the reverse shock photons on the forward shocked electrons
({\em dash-dotted curve}) and the forward shock photons on the reversely shocked
electrons ({\em dashed curve}). From Fig.1, it can be clearly seen that the SSC from 
the reverse shock dominates over the other three IC components
at gamma-ray bands less than  a few tens of GeV with { a peak around 
a few {\rm MeV}}. At  $\varepsilon\sim10-100{\,\rm MeV}$, the SSC component
of the reverse shock exceeds other IC components by about two orders of 
magnitude. { Although the peak location
of SSC emission strongly depends on this unknown parameter $xi_e$, this
emission dominates over other IC emissions for a wide range of shock
parameters.}

ii) In Fig.2, we present the energy spectra ($\nu{f_{\nu}^{IC}}$) of the IC emissions
with various shock parameters for the thin shell case. We find that a) for a wide
range of shock parameters, the SSC component from reverse shocks is the most 
important at  energy bands from tens of MeV to tens of GeV, to which EGRET is sensitive.
It dominates over both other IC  and the synchrotron emission components.
b) The SSC spectra from  reverse shocks above the turnover ($\nu>\nu_p^{IC}$) are
logarithmically more flattening than that of the seed synchrotron spectrum and varying
continuously . For small values of $p$, we even find that the IC energy output
peak well above $\min(\nu_c^{rs,IC}, \nu_m^{rs,IC})$, which is the peak emission
frequency for the approximate power-law IC spectrum. A similar result was reached by
Sari \& Esin (2001) for the SSC of afterglow emission.
{
Moreover, for small value of $p$ (e.g. $p=2.2$), the SSC emission from the reverse shock
dominates over the synchrotron and IC processes even in the TeV energy bands (see Fig. 2(d)).}
c) Fig.2 also suggest that strong TeV emission should also be emitted
from the two combined-IC  { and forward shock SSC  } processes  
for most GRBs.
{
For a moderate steep distribution of the shocked electrons (e.g. $p=2.5$), the combined-IC 
and/or forward shock SSC become increasing dominated at TeV bands.}
 However, it would only be detected from nearby, low-redshift
bursts for which the attenuation due to intergalactic infrared emission is small.

iii) EGRET has detected  prompt emission above $30{\rm MeV}$
 from several  bright GRBs
triggered by BATSE (Catelli et al. 1998), among which GeV 
photons have been detected from GRB930131 (Sommer et al. 1994; Ryan et al. 1994)
and GRB940217 (Hurley et al. 1994). 
{ 
GRB940217 even exhibits delayed GeV emission
90 minutes after the trigger. Several models have been proposed to explain
the delayed and prompt GeV emissions. For example,
M\'esz\'aros \& Rees (1994) proposed that the impact of a relativistic wind
from the central engine on the external matter may cause the delayed GeV emission;
 Katz (1994)
suggested that the impact of the fireball on a dense clouds could produce
high-energy gamma-ray emission via $\pi^0$ decay process;
Vietri (1997) and Totani (1998a,b) suggested that the synchrotron radiation
of the protons may be responsible for the GeV emissions; etc.
}
In a recent paper (WDL), we have suggested that the SSC emission from 
the reverse shock could explain both the flux level and the spectrum of the 
high energy gamma-rays detected by EGRET. We here compute the slope of the photon spectrum
at high energy bands and plot it in Fig. 3. We can see that at energy bands from
tens of MeV to tens of Gev, the photon spectrum index $\alpha$ (the photon number
$dn(h\nu)/d\nu\propto{\nu^{\alpha}}$) ranges from 1.7 to 2.15, which is consistent with
the observed high energy gamma-ray photon spectrum by EGRET from some bright
GRBs (e.g. Sommer et al. 1994).

iv) Then, we investigate the SSC emission from the reverse shock of a thick shell.
{
We compute the time-integrated SSC energy spectrum for
 a thick shell 
with typical parameters as $\Delta=3\times10^{12}$,  $E=10^{53}{\rm erg}$,
$p=2.5$ and $n_1=1$, but with different values of the initial Lorentz factor $\eta$:
$\eta=300$({\em dash-dotted curve}), $\eta=500$({\em dashed curve}) and $\eta=1000$ ({\em
dotted curve}). They are plotted in Fig. 4. in 
comparison  with the thin shell case ({\em solid curve}).
As $\eta$ increases, the peak frequency of the time-integrated SSC energy spectrum 
increase accordingly. The peak frequency of SSC emission of the thick shell 
is not far from that of the
thin shell case with the same $\eta$ and typical shock parameters,
though the reverse shock of the thick shell is relativistic. }
This is because that 
this IC emission peaks at $\nu_c^{rs,IC}$
 (this reverse shock is in the fast cooling regime), which is close to $\nu_m^{rs,IC}$
for the thin shell. Fig.4 also shows that 
the peak flux is much lower than the thin shell case. The reason is that 
at the peak time  of the reverse shock emission, the shell has traveled to a larger
distance, resulting a lower electron scattering optical depth in the shell.

\section{Summary and Discussion}
The fireball model for GRBs involving an ultrarelativistic fireball ejecta
expanding into an external medium has two successful  predictions: 
one is the afterglow emission
from the synchrotron process in the forward shock region (Katz 1994; M\'esz\'aros
\& Rees 1997), and another is the prediction of bright optical emission when
a reverse shock is present (Sari \& Piran 1999b). Subsequent observations of 
multiwavelength afterglows (e.g. Wijers, Rees \& M\'esz\'aros 1997) 
and a bright optical flash (Akerlof et al. 1999) gave basic confirmation of this
model. In this paper, we calculate the SSC and combined-IC emissions from these
two shock regions at very early phases.
For a wide range of shock parameters, the SSC emission from the reverse shock
dominates over the synchrotron and other IC emissions at energy bands from tens of MeV
to tens of GeV, while the combined-IC { and/or the forward shock SSC emissions }
become
 increasingly dominant at TeV energy bands.

We further compute the photon spectrum index $\alpha$ (viz. $dn(h\nu)/d\nu\propto
\nu^{-\alpha}$) of the reverse shock SSC emission, which
is $\alpha\sim2.0$ for typical shock
parameters at energy bands to which EGRET is sensitive.
 Based upon this and the SSC spectral flux level, we suggest that this
process can provide a plausible explanation for the prompt high energy gamma-rays
detected from some bright bursts, such as those from GRB930131, GRB910503 and
GRB940217 etc.  In  WDL, we also derived the decaying light curves of SSC
emission after the reverse shock peak time
and found that it decays quite rapidly, regardless
whether the observed band locates above or below the cooling break frequency
of the inverse Compton component.
The planned Gamma-ray Large Area Space Telescope (GLAST) mission will have
larger effective area and field of view than EGRET, and so will likely be
able to monitor the time evolution of the high energy gamma-ray flux.

At TeV energy bands, the combined-IC emissions and the SSC emission from the forward shock
become dominant over the SSC from the reverse shock for  intermediate values of
$p$ (e.g. $p=2.5$).
{ Nevertheless, for small values of $p$, the  SSC from the reverse shock
still dominates even in TeV energy bands. 
}
Recently the Milagro group reported evidence for TeV emission from GRB970417a, one
of the 54 BATSE GRBs in the field of view of their detector, Milagrito (Atkins et al.
2000). An excess of gamma-rays above background is seen during the durations of this
burst and the chance probability for detecting such an excess is estimated
to be less than $1.5\times10^{-3}$. 
{ 
Totani (2000) suggested that  
proton-synchrotron model of GRBs provides a possible explanation for these observational 
results. }
Our calculation is also consistent with detection
of TeV emission from GRBs that are near enough to avoid serious attenuation due to
intergalactic infrared radiation field.

{ 
Two currently popular models for GRBs are
the  mergers 
of compact objects (neutron stars or black holes) and the 
cataclysmic collapse of massive stars. In the former model, compact objects 
are expected to have significant spacial velocities so that their 
mergers would take place at many kiloparsecs outside their birthplaces. 
Thus, GRBs produced by this model would occur in the 
interstellar medium (ISM) with density $n\sim 1\,{\rm cm}^{-3}$.
However, if a collapsing massive star (Woosley 1993; Paczy\'{n}ski 1998) is the
origin of the GRBs, the circumburst medium is the wind ejected by the star prior
to its collapse, whose density decreases outwards. 
In the wind circumburst medium case, the external medium density is much higher than
the ISM case at the deceleration length-scale  and the magnetic field is much
higher accordingly. Therefore, from Eq.(2), we know that most of the shocked electrons 
cool to be sub or trans-relativistic ones ($\gamma_c-1\la{1}$) on the deceleration
 time $t_{dec}$ (also see Eq.(14) in Dai \& Lu 2001).
With such low values of $\gamma_c$, the self-absorbed
cyclo-synchrotron radiation and multiple Compton scatterings of these cooled
electrons may construct the low-energy part of the resulting spectrum. 
At the high-frequency part, the spectrum is similar to the fast-cooling spectrum of 
relativistic electrons, i.e.
the spectrum  is still in the form of $f_{\nu}\propto{\nu^{-1/2}}$
extending up to $\nu_m$ and then $f_{\nu}\propto{\nu^{-p}}$ upward,
produced by the cooling of newly-shocked electrons.  The resulting IC spectra 
in this case is therefore much more complicated. 
Also, we estimate that the IC spectra flux at GeV to TeV bands may be as high as
 that of the ISM case.
 }

{ We  wish to thank the anonymous referee for his/her constructive and careful comments
that enabled us to improve the manuscript.}
XYW also thanks Drs D. M. Wei and Z. Li for valuable discussions. 
This work was supported by the National Natural Science Foundation
of China under grants  19973003 and 19825109, and the 
National 973 project.

\newpage
\begin{center}
FIGURE CAPTION
\end{center}
{ Figure 1.} The spectra of the IC emissions at the reverse shock peak time
for typical shock parameters:
$E=10^{53}{\, \rm erg}$, $\xi_e=0.6$, $\xi_B=0.01$, $p=2.5$ and an ISM 
external medium with $n_1=1$. The {\em solid } and {\em dotted} curves represent the 
SSC emissions from the reverse shock and forward shock, respectively.
Also plotted are the IC emissions of scatterings of reverse shock photons
on the forward shock electrons ({\em dash-dotted curve}) and forward
shock photons on the reversely shocked electrons ({\em dashed curve}).\\
{~}\\
{ Figure 2.}The energy spectra of  synchrotron and IC emissions
  at the reverse shock peak time for the ISM circumburst environment case with various
  shock parameters: a)$E=10^{53}{\, \rm erg}$, $\xi_e=0.6$, $\xi_B=0.01$, $p=2.5$ 
  and $n_1=1$; b)$E=10^{52}{\, \rm erg}$, $\xi_e=0.6$, $\xi_B=0.01$, $p=2.5$ 
  and $n_1=1$;
  c)$E=10^{53}{\, \rm erg}$, $\xi_e=0.6$, $\xi_B=10^{-4}$, $p=2.5$ and $n_1=1$;
  d)$E=10^{53}{\, \rm erg}$, $\xi_e=0.6$, $\xi_B=0.01$, $p=2.2$ and $n_1=1$.
  The   {\em thin dash-dotted} and {\em dashed curves } represent the synchrotron
  spectra of the reverse shock and forward shock, respectively. The four IC spectra
  are shown by the curves in the same way as in Fig. 1.\\
{~}\\
{ Figure 3.} The high energy gamma-ray photon spectrum index $\alpha$ of the SSC 
emission from the reverse shock with shock parameters as used 
in Fig.1.\\ 
{~}\\
{ Figure 4.} Comparison of the time-integrated  
energy spectrum between the thin shell case ({\em solid curve}) 
with shock parameters as used in Fig.1 
and 
{ thick shell case  with the same shock parameters $E=10^{53}{\, \rm erg}$,
 $\Delta=3\times10^{12}\, {\rm cm}$, $p=2.5$ and $n_1=1$ but with
 different values of $\eta$ : $\eta=300$({\em dash-dotted curve}),
  $\eta=500$({\em dashed curve}) and $\eta=1000$ ({\em
dotted curve}).
}\\

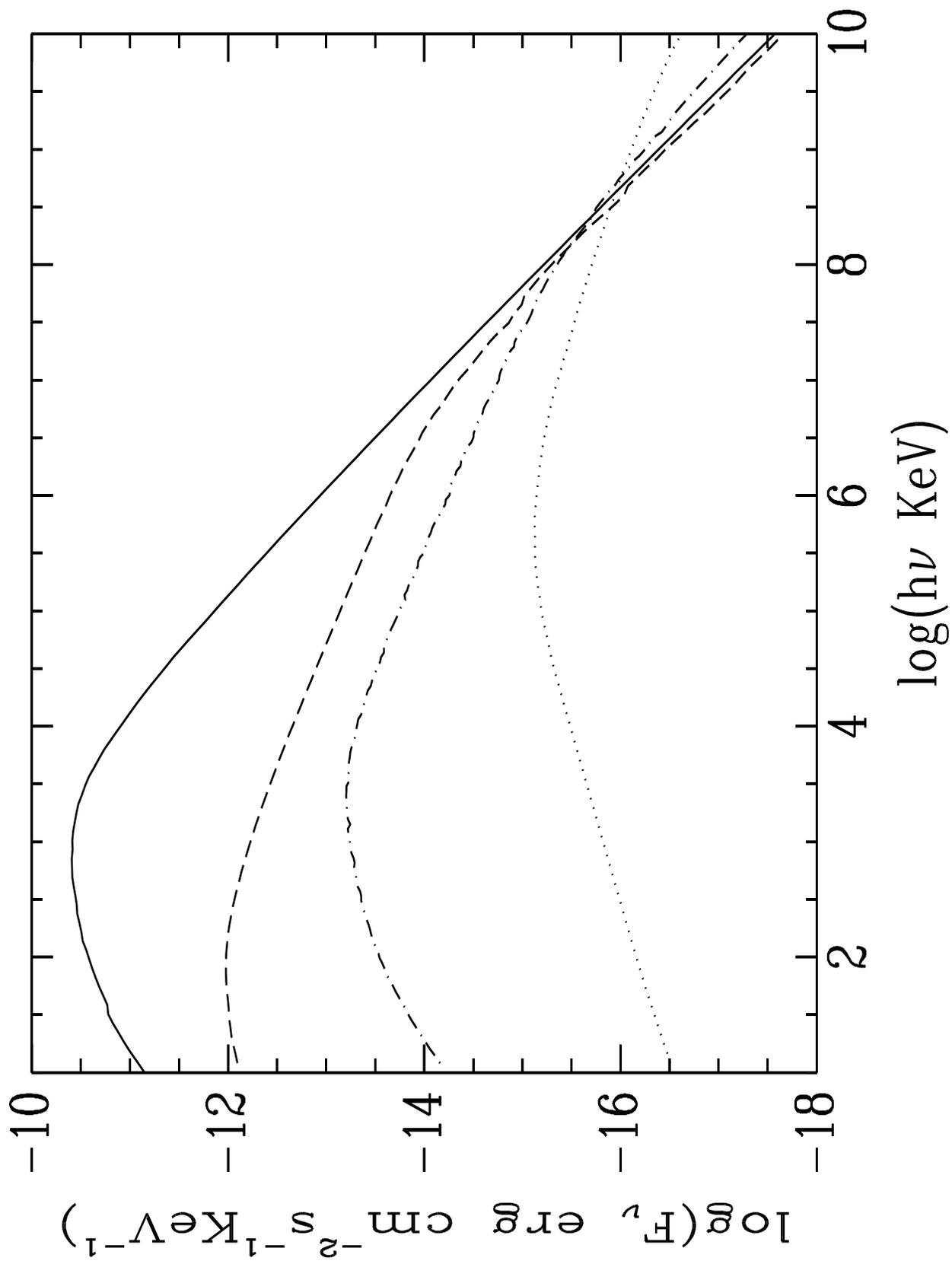
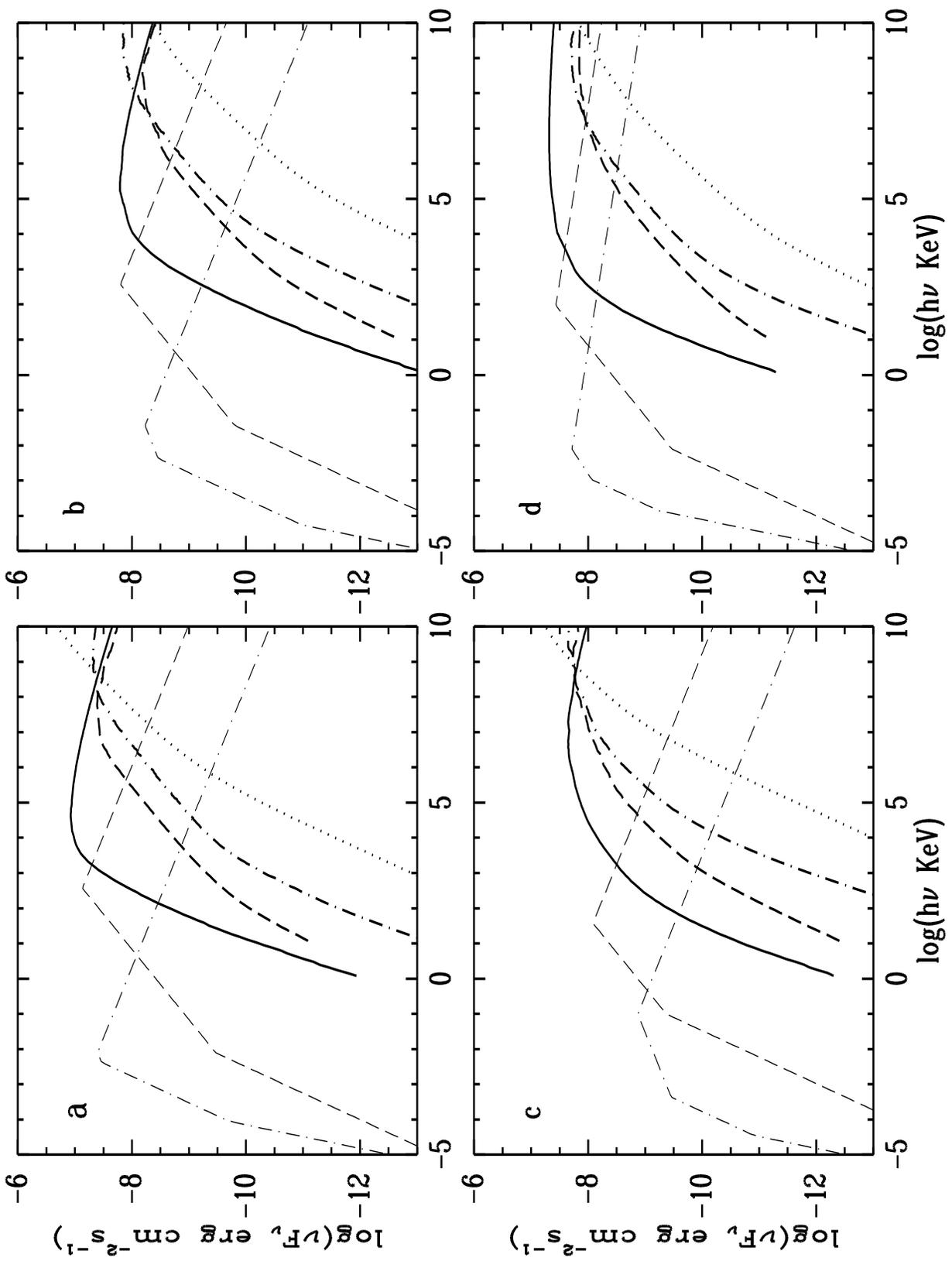
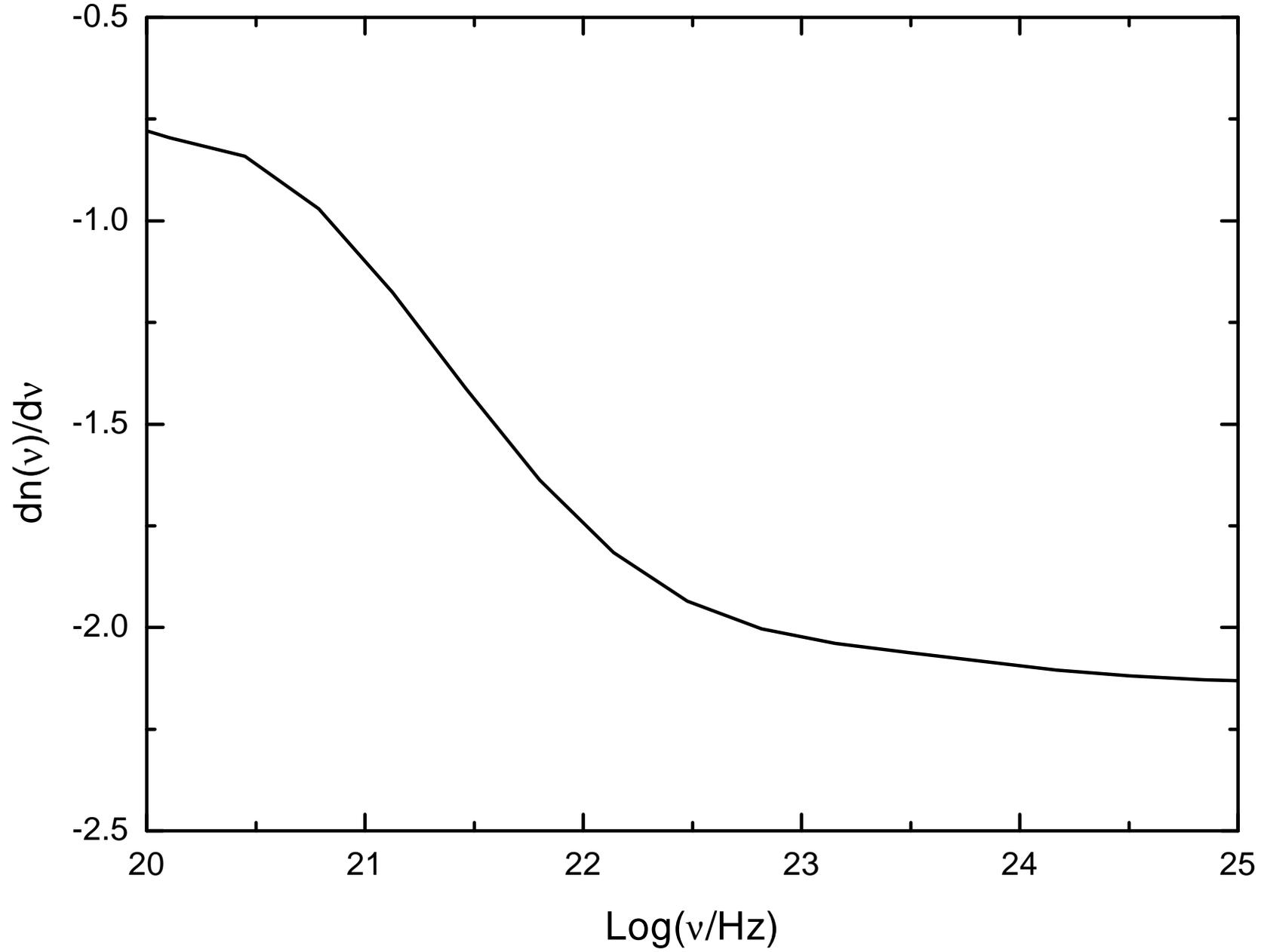
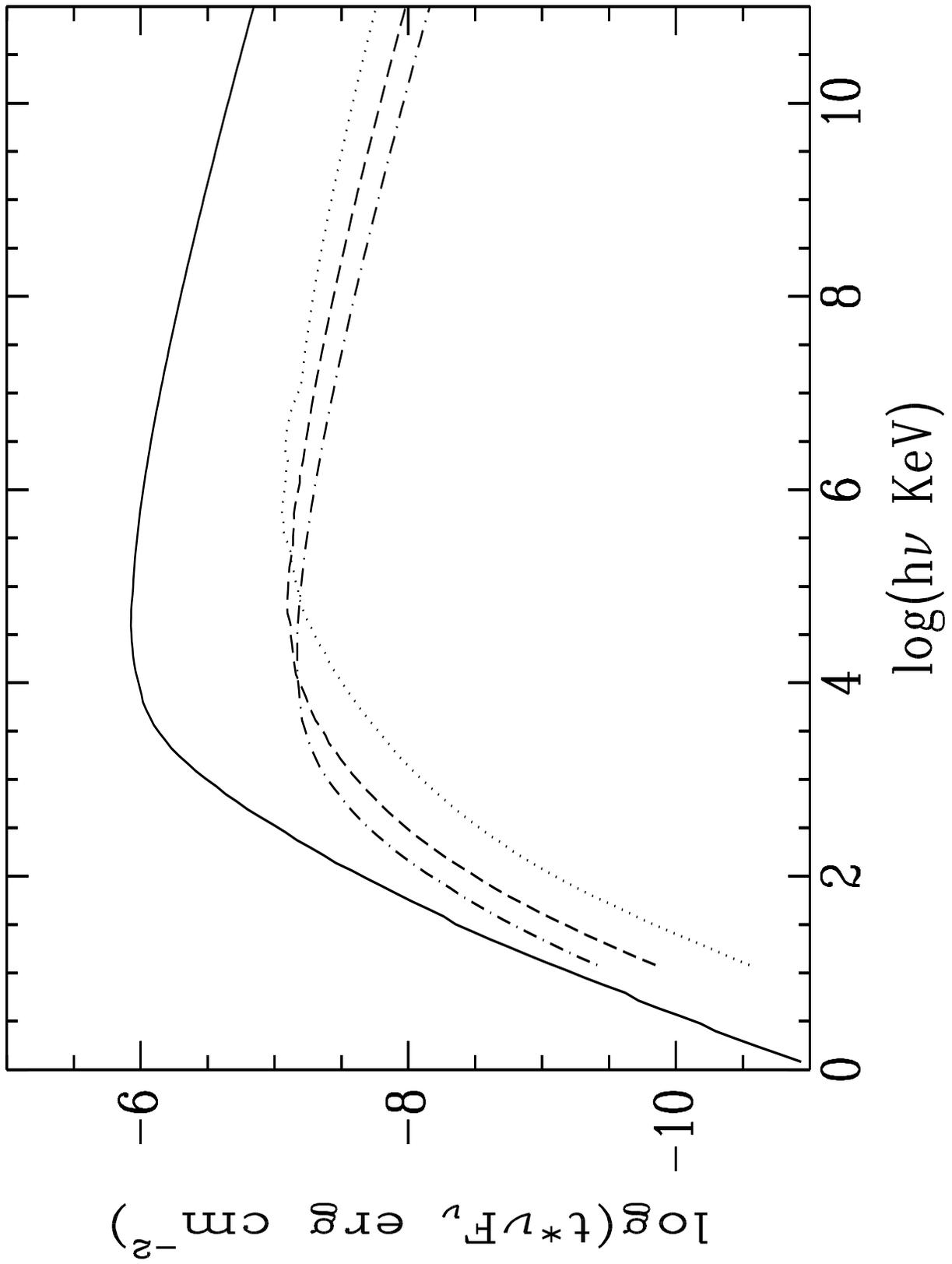
\end{document}